\documentclass[aip,amsmath,amssymb,reprint,groupedaddress]{revtex4-1} \makeatletter 

\usepackage{graphicx}
\usepackage{dcolumn}
\usepackage{bm}

\usepackage[utf8]{inputenc}
\usepackage[T1]{fontenc}
\usepackage{xurl} 
\usepackage{mathptmx}
\usepackage{etoolbox} 

\usepackage{hyperref} 

\usepackage{xcolor}
\hypersetup{
    colorlinks,
    linkcolor={red!50!black},
    citecolor={green!50!black},
    urlcolor={blue!70!black} 
}
\usepackage{tabularx,booktabs}

\def\@email#1#2{%
 \endgroup
 \patchcmd{\titleblock@produce}
  {\frontmatter@RRAPformat}
  {\frontmatter@RRAPformat{\produce@RRAP{*#1\href{mailto:#2}{#2}}}\frontmatter@RRAPformat}
  {}{}
}%
\makeatother

\begin{document} 
 \begin{flushright} \includegraphics[width=4cm]{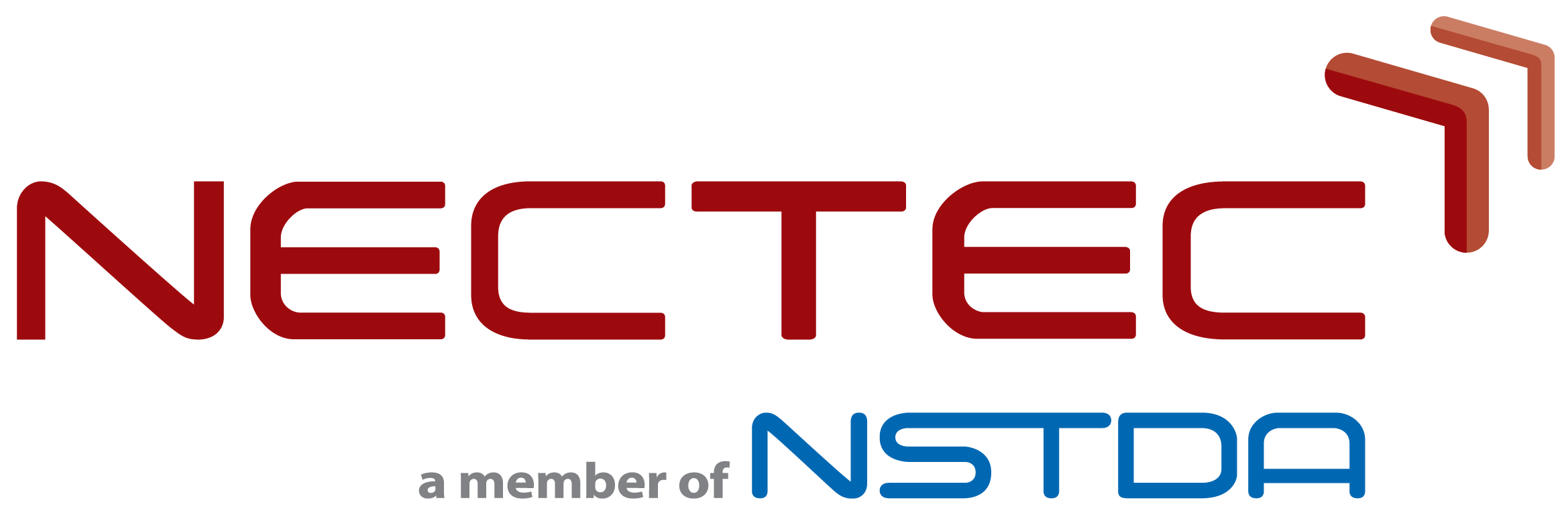}   \end{flushright}  

\vspace{-4mm}
\title{ Multiplexing quantum tunneling diodes for random number generation\vspace{2mm}}

\author{Kanin Aungskunsiri$^{*,\dagger}$} 


\author{Ratthasart Amarit$^\dagger$}

\author{Sakdinan Jantarachote$^\dagger$}

\author{Kruawan Wongpanya} 
\author{Pongpun Punpetch} 
\author{Sarun Sumriddetchkajorn} 


\affiliation{ \vspace{1mm}
National Electronics and Computer Technology Center,\\
National Science and Technology Development Agency,\\
Ministry of Higher Education, Science, Research, and Innovation,
\mbox{112 Thailand Science Park, Phahonyothin Road, Khlong Nueng, Khlong Luang, Pathum Thani, 12120 Thailand}}

\date{\today , \href{https://arxiv.org/abs/2212.12177}{arXiv:2212.12177}} 



\begin{abstract}


\normalsize
\vspace{-3mm} \noindent $^*$The author to whom correspondence may be addressed: \href{https://orcid.org/0000-0002-9333-7440}{\textit{kanin.aungskunsiri@nectec.or.th}}
\normalsize 

\vspace{1mm}\noindent $^\dagger$These three authors contributed equally to this work.\vspace{4mm}
\normalsize

\noindent\vspace{1mm}\textbf{ABSTRACT} \vspace{2mm}\\
\noindent Random numbers are indispensable resources for application in modern science and technology. Therefore, a dedicate entropy source is essential, particularly cryptographic tasks and modern applications. In this work, we experimentally demonstrated a scheme to generate random numbers by multiplexing eight tunnel diodes onto a single circuit. As a result, the data rate of random number generation was significantly enhanced eightfold.  In comparison to the original scheme that employed one diode, this multiplexing scheme produced data with higher entropy. These data were then post-processed with the Toeplitz-hashing extractor, yielding final outputs that achieved almost full entropy and satisfied the U.S. National Institute of Standards and Technology (NIST) Special Publication 800-90B
validation. These data also passed the NIST Special Publication 800-22 statistical randomness examination and had no sign of patterns detected from an autocorrelation analysis. 

\vspace{2mm}\noindent \small  
\textbf{Note:} 
\textit{This article may be downloaded for personal use only. Any other use requires prior permission of the author and AIP Publishing. This article appeared in \href{https://doi.org/10.1063/5.0113995}{Rev. Sci. Instrum. \textbf{94}, 014704 (2023)} and may be found at \href{https://doi.org/10.1063/5.0113995}{https://doi.org/10.1063/5.0113995}.}

\normalsize

\end{abstract}

\maketitle

\section{Introduction}


When discussing random numbers, one may think of lotteries and drawing lots. In fact, random numbers are used widespread on a daily basis. These numbers are employed by mobile phones for encryption, authentication, and communication. Nowadays, digital banking services and web applications employ random numbers for the generation of one-time passwords used for authentication and transaction verification \cite{Young_Sil2010}. Applications of random numbers have been found in scientific simulation \cite{Metropolis1949}, weather forecasting, and cryptography \cite{Gennaro2006, Shannon1949}. In social sciences, random numbers are resources used for sampling a random dataset that represents an entire population used in conducting surveys \cite{Lohr2010}. Cryptographic network protocols over the internet, such as the Secure Shell (SSH), the Secure Sockets Layer (SSL) \cite{Weaver2006}, and the Transport Layer Security (TLS), employ random numbers to securely operate network services over an untrusted networks. Upcoming quantum technologies require random number generators to setup private keys used in quantum key distribution \cite{BB84, Ekert1991, Zhang2014} and blind quantum computing \cite{Fitzsimons2017}.

A random number generator may be derived from a computerized algorithm to produce random sequences with high throughput. This generator implements a deterministic process that converts a small input string, as a seed, into a larger set of strings that appears to be sufficiently random. However, the outcome may be predicted if the seed and algorithm are known. Alternatively, random number generation, in association with natural phenomena, such as radioactive decay \cite{Alkassar2005, Park2020}, chaotic process \cite{Argyris2010, Gleeson2002, Reidler2009, Uchida2008, Wishon2018}, atmospheric noise \cite{Marangon2014}, cosmic background radiation \cite{Lee2017}, and quantum mechanical properties of light \cite{Applegate2015, Jennewein2000, Nie2015, Nie2014, Rarity1994, Shi2016, Symul2011, Wahl2011}, is a solution to generating truly random number generator.

Our previous work \cite{Aungskunsiri2021} demonstrated the application of a quantum tunnelling diode to realize an entropy source. The scheme harnessed inexpensive electronics and did not require a complex system for implementation. In this work, we extended the capability of the original scheme by multiplexing multiple tunnel diodes onto a single circuit. The detailed method and experimental results are discussed in the following parts of this paper.


\begin{figure}[!t]
  \includegraphics[width=8.0cm,keepaspectratio]{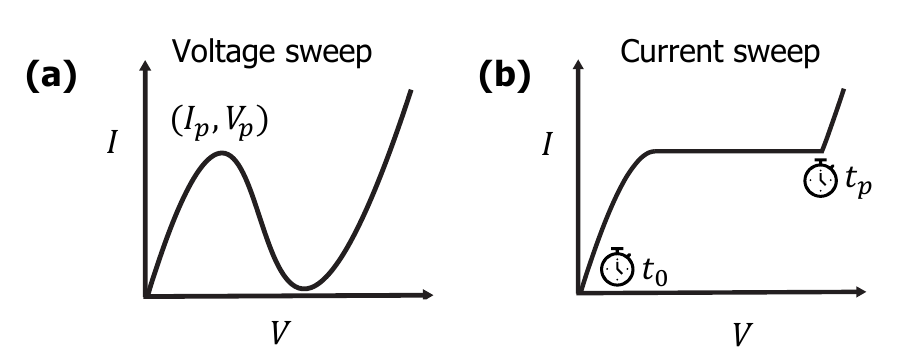}
    \caption{(a) Ideal I-V characteristic of a tunnel diode from a voltage sweep. (b) Forward current sweeping causes a sudden change of voltage when current reaches a peak value ($I_p$) at an unfixed time interval ($t_p$).}
\end{figure}

\section{Method}

A tunnel diode \cite{Esaki1958} is a solid-state device incorporated with a p-n junction. Because this device was made to feature a thin depletion region, current flow at low biased voltage is due to the collective behavior of tunneling electrons. By applying a voltage sweep across the device, an ideal plot of current ($I$) as a function of biased voltage ($V$) features an N-shaped characteristic (Figure 1a). On the other hand, applying a current sweep causes an immediate change in the voltage at peak point ($I_p,V_p$), which is unfixed \cite{Bernardo2017} and has statistical fluctuation described by a normal distribution \cite{Aungskunsiri2021}. 

According to our previous work \cite{Aungskunsiri2021}, the values of $I_p$ vary on a scale of microamperes. During current sweeping, taking an I-V measurement to obtain precise values of $I_p$ is difficult. This process may require a high-end equipment, such as a current-source meter, to sweep current and collect a dataset of I-V values. Because the dataset acquired from this method is discrete-time signals, detecting the values of $I_p$ with a resolution of sub-microampere range would be a time-consuming form of data acquisition. Accordingly, taking direct measurement of $I_p$ values for random number generation would be impractical with limited data rates at several kbit/s. Alternatively, measuring time interval, $t_p$, when the voltage has a sudden jump is simpler with the use of a time-counting module (Figure 1b). Because a value of $t_p$ is in association with the event of voltage jump at a peak current ($I_p$), measuring a value of $t_p$ can also refer to the associated value of $I_p$; hence, an entropy source can also be realized from the measured values of $t_p$.

\begin{figure}[!htbp]
  \includegraphics[width=7.0cm,keepaspectratio]{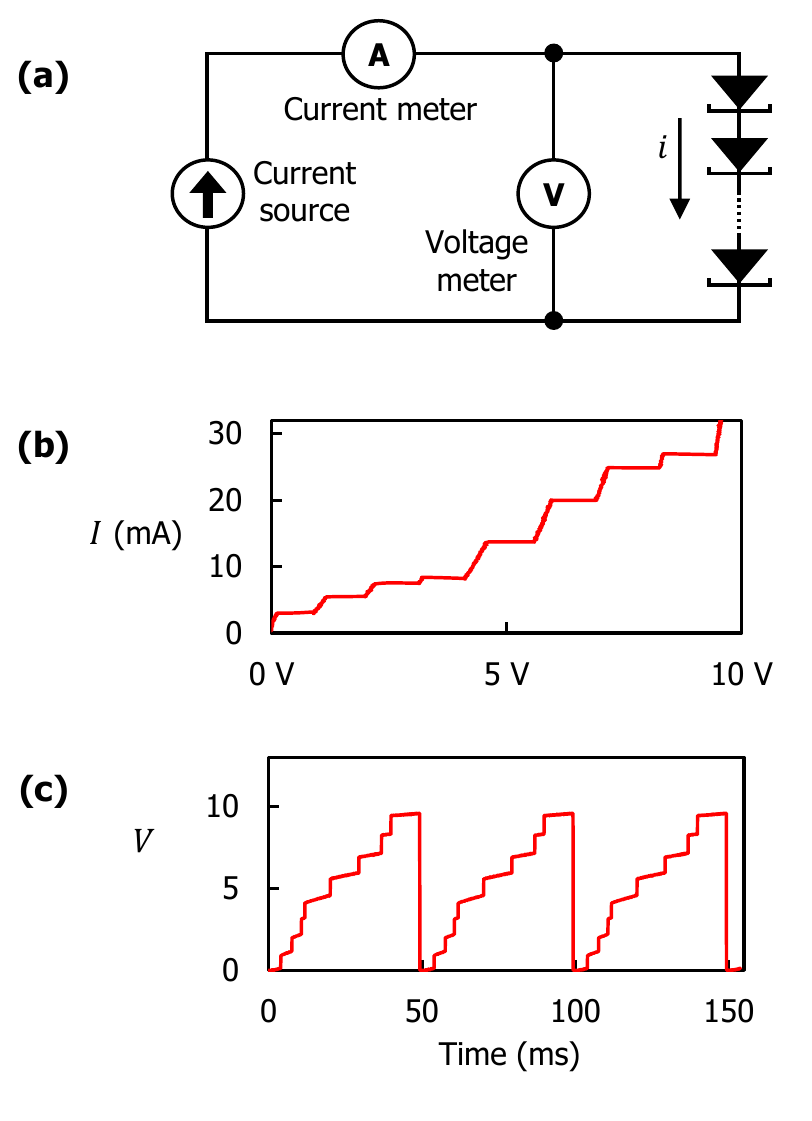}
  \vspace{-2mm}
    \caption{(a) Circuit diagram of the experiment. (b) I-V plot according to one current sweep. (c) Time series of voltage signals obtained from multiple current sweeps. In (b) and (c) sawtooth waves with a 20 Hz repetition rate were applied for current sweeping.}
\end{figure}


\begin{table}[!ht]

\centering 
\caption{Tunnel diodes used in this work.}\vspace{0.2cm}

\begin{ruledtabular}
\begin{tabular}{cccc}

Diode      & Part number & $I_p$ & $V_p$ \\ \cmidrule{1-4}
D1	& 3I306E	& 2.7 mA	& 1.21 V \\
D2	& AI301V	& 4.1 mA	& 1.25 V \\
D3	& 3I101I	& 5.1 mA	& 1.25 V \\
D4	& 3I201A	& 8.5 mA	& 1.20 V \\
D5	& 3I201D	& 12.3 mA	& 1.27 V \\
D6	& 3I201D	& 18.9 mA	& 1.16 V \\
D7	& 3I201D	& 24.3 mA	& 1.10 V \\
D8	& 3I201D	& 25.6 mA	& 1.24 V \\

\end{tabular}
\end{ruledtabular}
\end{table}

This work extended the capability to realize an entropy source from a single diode to multiple diodes connected in a series (Figure 2a). The key idea relies on the selection of individual diodes with different values of $I_p$ giving that a current sweep produces multiple voltage jumps at different time intervals. It is essential that all $I_p$ values have clear separation so that all voltage jumps are distinguishable; otherwise it would be difficult to differentiate multiple voltage jumps that are in association with overlapped $I_p$ values and data acquisition from a current sweep may not produce a complete set of desired data. To demonstrate this concept, eight tunnel diodes (D1–D8) from various part numbers as listed in Table I were chosen for implementation. These diodes feature different values of $I_p$, providing that the variations of $I_p$ do not overlap. Here, four tunnel diodes (D5–D8) were issued with the same manufacturer part number, though they still had variations in $I_p$ as a result of manufacturing errors. In a series circuit, the amount of current is the same through any component, the voltage supply is equal to the sum of the individual voltage drops, and the order arrangement of tunnel diodes in this scheme is negligible.

By performing a forward current sweep, we obtained an I-V plot, as shown in Figure 2b. There were eight voltage jumps corresponding to moments that the sweeping current reached $I_p$ for each of individual diodes. An example for voltage time series is presented in Figure 2c. To obtain a dataset of $t_p$, we followed the flow diagram, as presented in Figure 3, and implemented a circuit diagram, as presented Figure 4. The physical implementation is illustrated in Figure 5.


\begin{figure*}[!htbp]
  \includegraphics[width=16cm,keepaspectratio]{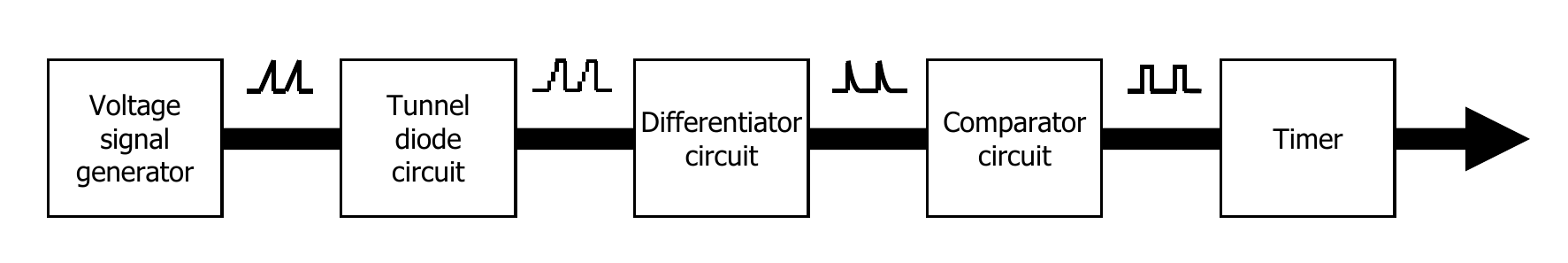}
  \vspace{-2mm}
    \caption{Flow diagram of the experiment. A waveform generator (Hantek HDG2022B) was configured to produce sawtooth voltage signals with a 10 kHz repetition rate. The voltage signals were then converted to current source used for current sweeping at the tunnel diode circuit.}
\end{figure*}

\begin{figure*}[!htbp]
  \includegraphics[width=15cm,keepaspectratio]{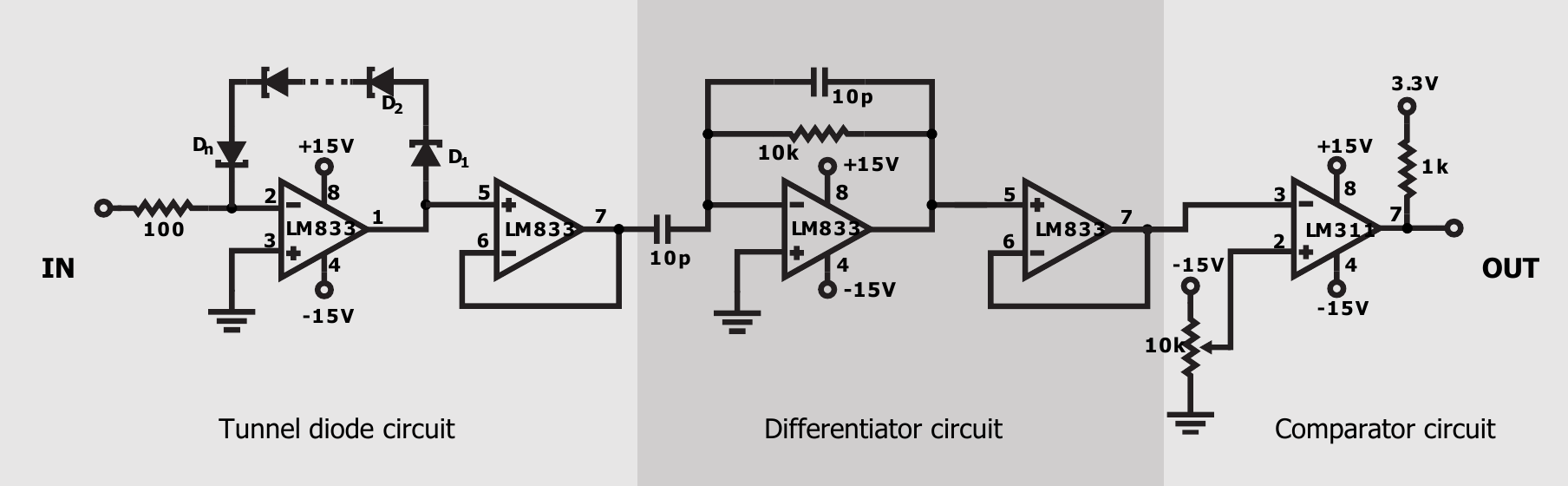}
  \vspace{-2mm}
    \caption{Circuit diagram implemented in this work. At the input, sawtooth waves produced from a waveform generator were fed into the tunnel diode circuit.  At the output, eight transistor–transistor logic (TTL) pulses were generated per single current sweep.}
\end{figure*}

\begin{figure*}[!htbp]
\includegraphics[width=15cm,keepaspectratio]{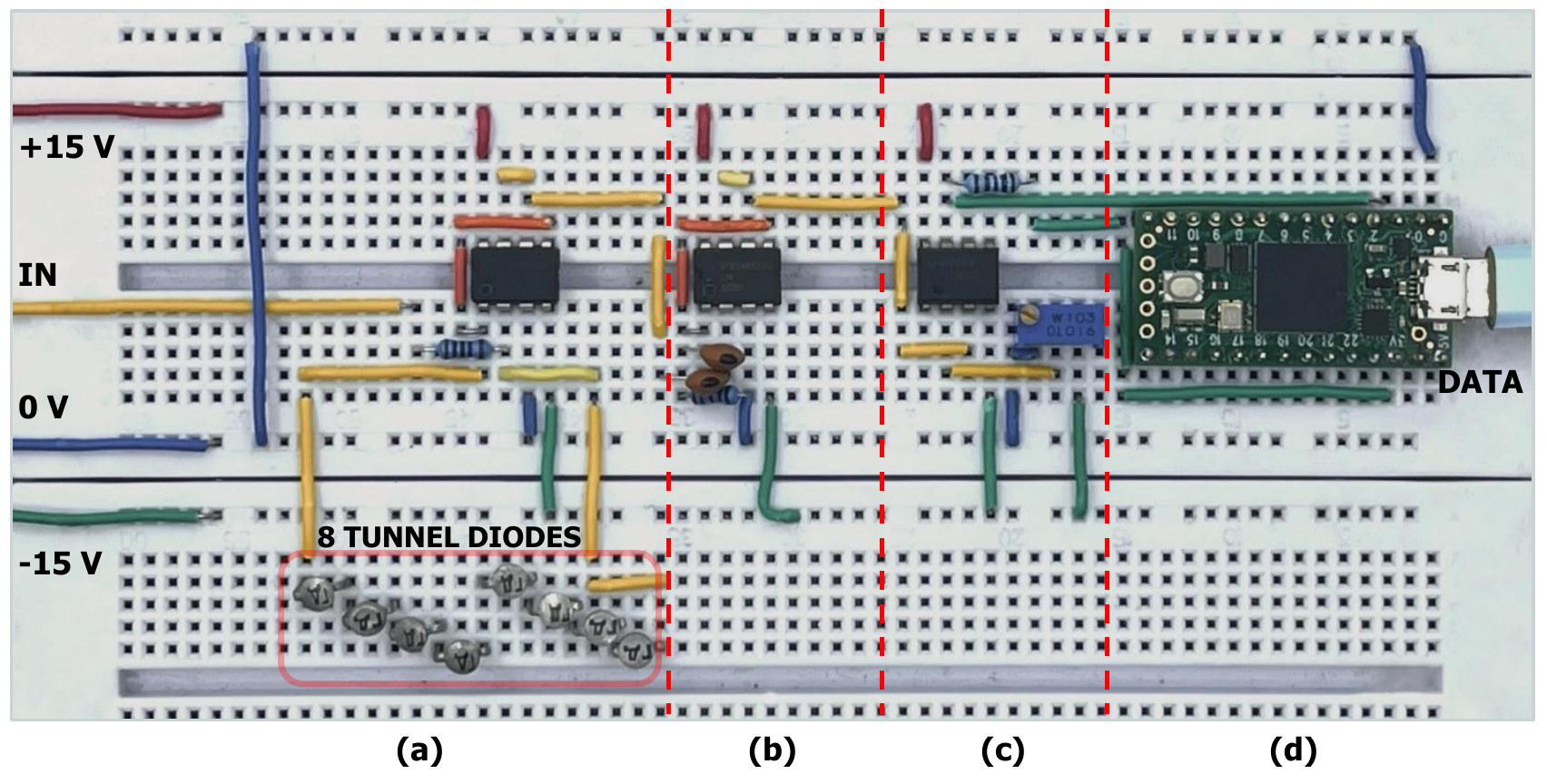}
\vspace{-2mm}
\caption{Experimental setup with eight tunnel diodes. The setup consisted of (a) a tunnel diode circuit, (b) a differentiator circuit, (c) a comparator circuit, and (d) a timer. The TEENSY$\textsuperscript{\textregistered}$ Development Board 4.0 was employed as the timer.}   
\end{figure*}

Here, we discuss details of the multiplexing scheme for each stage of implementation. For ease of illustration, ideal outputs corresponding to the implementation with four diodes are presented in Figure 6. Sawtooth waves with a 10 kHz repetition rate were applied for current sweeping (Figure 6a). The waveform's amplitude was set at 30 mA, which is beyond all possible values of $I_p$ listed in Table I. Taking one current sweep, the time series of the voltage sum measured across all diodes in the series circuit features multiple voltage jumps at different time intervals (Figure 6b). The voltage signal was sent to a differentiator circuit to detect rates of voltage change and converted into a series of multiple pulses in association with the events of voltage jumps (Figure 6c). These pulses were passed through a comparator circuit for converting into two-level signals (Figure 6d). 
The TEENSY$\textsuperscript{\textregistered}$ Development Board 4.0, as a time-counting module, was used to obtain time intervals of two-level signals, from which a series of $t_p$, as \textit{raw signals}, were obtained (Figure 6e) with an accuracy of 1.18 ns.

\begin{figure}[!htbp]
  \includegraphics[width=7.0cm,keepaspectratio]{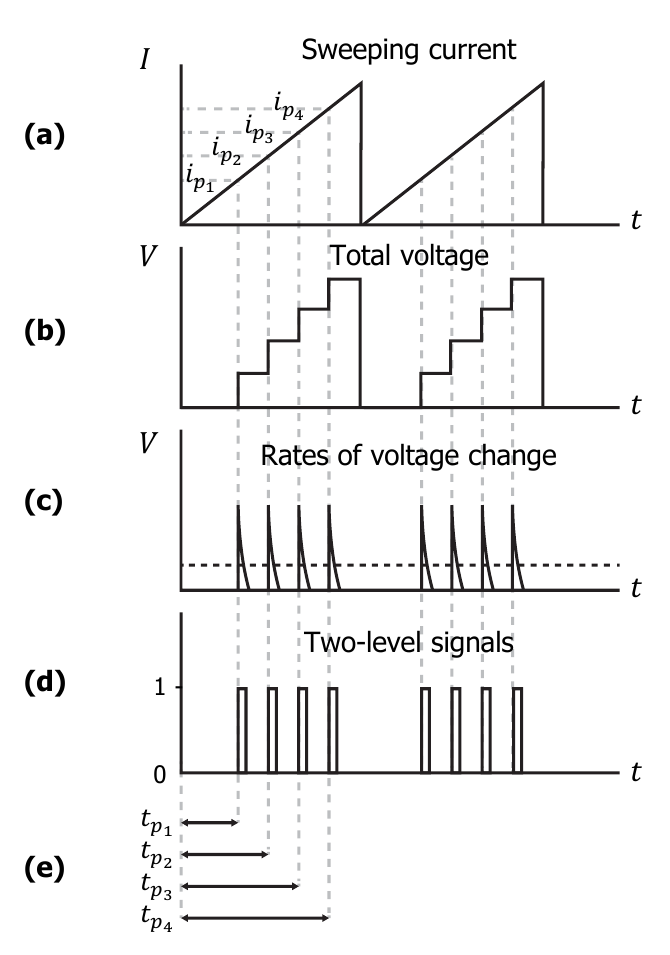}
  \vspace{-2mm}
    \caption{Ideal output signals from each stage of implementation. (a) Sweeping current applied to the tunnel diode circuit. (b) Voltage signals measured across the tunnel diode circuit. (c) Voltage signals in association with rates of voltage change acquired from a differentiator circuit and a threshold voltage (dashed horizontal line) for digital conversion. (d) Series of two-level signals obtained from a comparator circuit. (e) Series of $t_p$ obtained from a timer.}
\end{figure}

\section{Random bit extraction}


\begin{figure}[!htbp]
  \includegraphics[width=8.0cm,keepaspectratio]{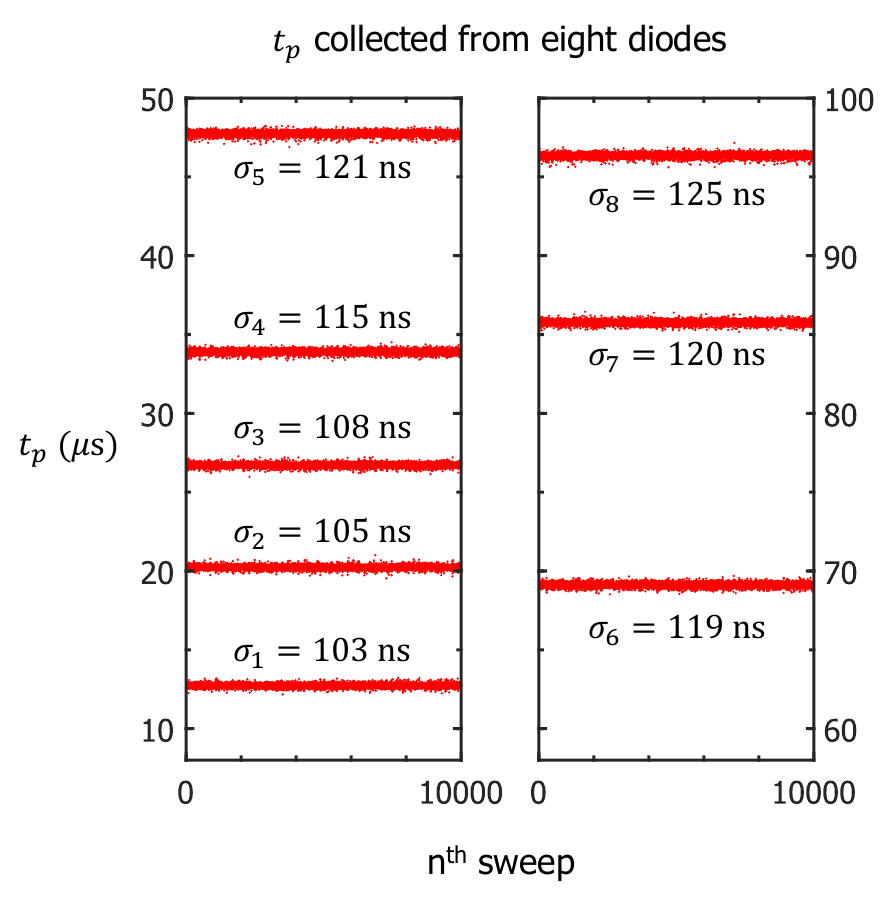}
  \vspace{-2mm}
    \caption{Time series of $t_p$ values obtained from eight tunnel diodes. Standard deviations ($\sigma_1, \sigma_2,.., \sigma_8$) were calculated from data collected for 10,000 current sweeps.}        
\end{figure}

\begin{figure*}[!htbp]
\includegraphics[width=14cm,keepaspectratio]{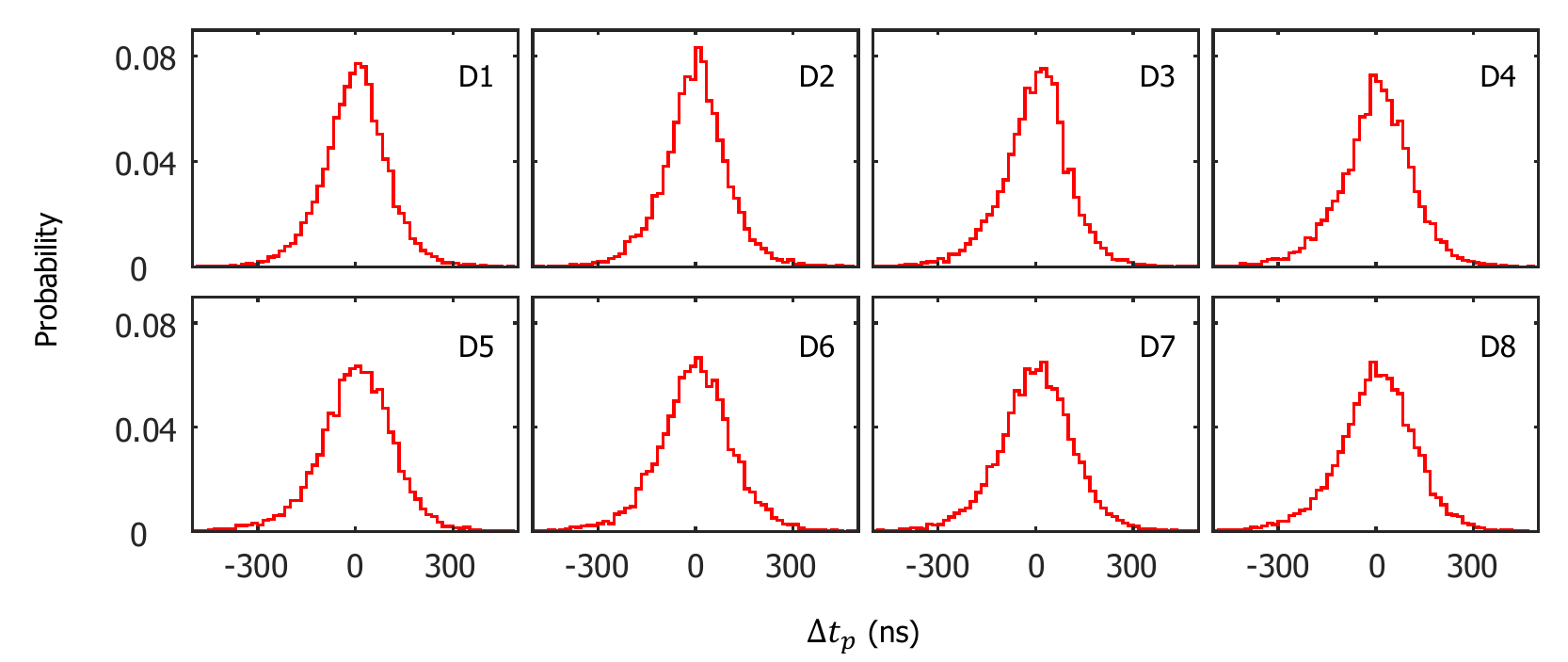}
\caption{Histograms of raw signals obtained from eight tunnel diodes (D1–D8). The value of $\Delta t_p$ at the horizontal axis corresponds to the deviation of the value of $t_p$ from its mean value. Data were collected from 10,000 current sweeps. The histograms have a bin size of 16.7 ns.}
\end{figure*}


\begin{table}[!t]
\centering

\caption{Statistical values of $t_p$ obtained from data collected from 10,000 current sweeps. Assuming normal distribution, the width, $w$, that covers 99.7\% of data is estimated from six standard deviations of $t_p$ values. Because the timer used in this work has a resolution of 1.67 ns, each data point fluctuates within $m\approx$ ceiling($w$/1.67) discrete values, where $m$ is the number of possible values. Maximum number of bits, $N$, is estimated from floor$[\log_2(w/1.67)]$.}

\vspace{0.2cm}

\begin{ruledtabular}
\begin{tabular}{ccccc}

 & Mean value   &	 Standard deviation &	$m$  &	$N$\\   \cmidrule{1-5}        
$t_{p_1}$	& 12,738 ns &	102.97 ns &	370 values &	8 bits  \\
$t_{p_2}$	& 20,236 ns &	105.27 ns &	379 values &	8 bits  \\
$t_{p_3}$	& 26,721 ns &	108.02 ns &	389 values &	8 bits  \\
$t_{p_4}$	& 33,910 ns &	115.41 ns &	415 values &	8 bits  \\
$t_{p_5}$	& 47,747 ns &	121.35 ns &	436 values &	8 bits  \\
$t_{p_6}$	& 69,125 ns &	118.56 ns &	426 values &	8 bits  \\
$t_{p_7}$	& 85,767 ns &	119.86 ns &	431 values &	8 bits  \\
$t_{p_8}$	& 96,358 ns &	125.11 ns &	450 values &	8 bits  \\

\end{tabular}
\end{ruledtabular}

\end{table}

Figure 7 presents time series of raw signals collected from 10,000 current sweeps. In each current sweep, we obtained eight data points that fluctuated at different ranges of time intervals. See Figure 8 for histograms collected from each diode. The statistical values of $t_p$, according to the experimental data, is summarized in Table II. For ease of interpretation, values are presented in the unit of nanosecond. In fact, data were originally taken in the form of counter unit that was in sync with the micro-controller clock and had a discrete integer value corresponding to a time interval of 1.67 ns.

Data in the form of counter unit were used for the random bit extraction. Taking a value of $t_p$ that has a discrete distribution with $m$ time bins; this data point was then converted into an integer $X$, for $X \in [0,2^N-1]$ and $m>2^N$, such that
\begin{equation}
X = t_p\  \text{modulo}\  2^N.
\end{equation}
Subsequently, $X$ was then digitized into a data of $N$-bit sequences.

According to Table II, each value of $t_p$ was digitized into a data of 8-bit sequences, as \textit{raw data}. We obtained a data of 64-bit sequences per one current sweep. By implementing bulk current sweeps with a repetition rate of 10 kHz, a data rate of 640 kbit/s could be obtained.

\section{Statistical analyses}

We then presented an entropy analysis of the noise source. In information theory, entropy of a dataset is a statical measure of randomness. A dataset of random bit strings with uniformly probability distribution has a maximum entropy of 1 per bit. A particular quantity known as \textit{min-entropy}, $H_\infty$, is the worst-case bound assessed from various entropy estimations. This quantity reflects the probability of guessing the correct output at the first trial from the noise source that is calculated as $2^{-H_\infty}$.

The process of entropy evaluation involves the property being independent and identically distributed (IID). A dataset is IID if, and only if, each sample in the dataset is mutually independent and has the same probability distribution. The U.S. National Institute of Standards and Technology (NIST) Special Publication (SP) 800-90B \cite{NIST_800_90B} suggested a methodology to assess whether the dataset is IID or not. Entropy evaluation depends on the IID test result; if the dataset is IID, entropy is calculated from the most common value (MCV) estimation. Otherwise, the assessed entropy is taken from the lowest value of the entropy calculated from the MCV estimation and 9 additional estimators \cite{90B}.

We collected raw data from 30,000,000 current sweeps. These data, labelled R0, contained 64-bit sequences per block, with each block corresponding to a series of raw data extracted from eight diodes in one current sweep. These data were separated into eight subsets of data, labelled R1 to R8, with each subset corresponding to raw data extracted from one individual diode. Taking R0 to R8, these data were split into five datasets of equal size. We implemented a software \cite{software} provided by the NIST with these data to run an IID check and perform an entropy evaluation. The result indicates that all datasets failed IID validation. Table III presents the assessed values of min-entropy according to the NIST SP900-90B \cite{NIST_800_90B}. As the table presents, datasets collected from all diodes (R0) had higher entropy than those collected from individual diodes (R1 – R8). Based on this result, it may be argued that raw data obtained from this multiplexing scheme were more random than those obtained from the original scheme that employed one diode.

If random variables are collected from multiple independent noise sources, a collection of these data is more random than data taken from each individual, which explains why a dataset obtained from eight diodes had higher entropy than subsets of data that corresponds to individual diodes. Based on this result, it is still inconclusive as to whether each of the individual diodes was considered an independent noise source. It is arguable that there might be correlations among random numbers generated from different diodes in one current sweep. However, this argument is negligible because we consider the entire system with eight diodes a single noise source and randomness of data at the final stage is of the most interest. 


\begin{table}[!htb]

\centering

\caption{Non-IID entropy evaluation according to the NIST SP800-90B. Each dataset corresponds to raw data collected from 2,000,000 current sweeps. Data labelled R0 correspond to raw data collected from eight diodes arranged in sequence. Data labelled R1–R8 corresponded with raw data extracted from individual diodes.}

\vspace{0.2cm}

\begin{ruledtabular}
\begin{tabular}{ccccccc}

& \multicolumn{6}{c}{Assessed min-entropy per bit} \\ \cmidrule{2-7}
 &	Dataset 1 &	Dataset 2 &	Dataset 3 &	Dataset 4 &	Dataset 5 &	Average\\ \cmidrule{1-7}
R1 &	0.5145 &	0.5042 &	0.4949 &	0.5060 &	0.5027 &	0.5045 \\
R2 &	0.5055 &	0.4691 &	0.4917 &	0.5018 &	0.4989 &	0.4934 \\
R3 &	0.5361 &	0.5312 &	0.5240 &	0.5266 &	0.5257 &	0.5287 \\
R4 &	0.5382 &	0.5118 &	0.5312 &	0.5380 &	0.5291 &	0.5297 \\
R5 &	0.5623 &	0.5666 &	0.5680 &	0.5725 &	0.5659 &	0.5670 \\
R6 &	0.5943 &	0.6019 &	0.6026 &	0.6037 &	0.5972 &	0.6000 \\
R7 &	0.5757 &	0.5741 &	0.5781 &	0.5780 &	0.5707 &	0.5753 \\
R8 &	0.5617 &	0.5613 &	0.5640 &	0.5669 &	0.5643 &	0.5637 \\ \cmidrule{1-7}
R0 &	0.8257 &	0.7427 &	0.8178 &	0.7513 &	0.7218 &	0.7718 \\

\end{tabular}
\end{ruledtabular}
\end{table}

This paper now focuses on raw data taken from the entire system. According to the entropy analysis, these data contain measurable biases and were not applicable for real-world use. The raw data were submitted to the Toeplitz-hashing extractor \cite{Ma2013, Zheng2019} for quality improvement and rendering IID outputs. This extractor converts an $n$-bit input data into an $m$-bit output data such that
\begin{equation}
\text{[Out]}_{m \times 1} = T_{m\times n}\cdot \text{[In]}_{n \times 1},
\end{equation}
where $T$, as a Toeplitz matrix, is a random binary matrix, though the diagonal elements are fixed to 1. The dimensions of the Toeplitz matrix satisfy $m \leq  nH_{\infty}$, where $H_{\infty}$ is the entropy rate of data. A statical distance between the post-processed data and ideal random data is quantified with a security parameter, $\epsilon$, which is calculated as $2^{(m-nH_\infty)/2}$.

For simplicity of implementation, $n$ and $m$ were chosen to be multiples of 8 (bit length for one byte). Given that $H_\infty$ of the raw data is 0.7718, and $n$ = 2048, we chose $m$ = 1320 to obtain $\epsilon < 2^{-100}$. After post-processing, the bit length of the post-processed data, as \textit{final outputs}, was reduced by 35.55\%.

The final outputs were re-submitted for entropy analysis with the NIST SP800-90B \cite{NIST_800_90B}. This time, the derived data passed IID validation, suggesting that entropy assessment was determined from the MCV estimation. The result, as presented in Table IV, shows a quality improvement with an approximate average min-entropy of 0.9997 per bit.

\begin{table}[!htb]

\centering
\caption{Assessed min-entropy of raw data versus final outputs. The final outputs in each dataset were derived from 247,500,000-bit sequences of raw data.} 
\vspace{0.2cm}

\begin{ruledtabular}
\begin{tabular}{lccccc}

& \multicolumn{5}{c}{Assessed min-entropy per bit} \\ \cmidrule{2-6} 
 & Dataset 1 & Dataset 2 & Dataset 3 & Dataset 4 & Dataset 5 \\ \cmidrule{1-6} 
Raw data &	0.8257 &	0.7427 &	0.8178 &	0.7513 &	0.7218 \\
Final outputs &	0.9997 &	0.9997 &	0.9997 &	0.9996 &	0.9998 \\

\end{tabular}
\end{ruledtabular}
\end{table}

We then aimed to validate the statistical randomness of these final outputs. Thousand datasets of 1,000,000-bit sequences was examined with 15 methods in accordance with the NIST SP800-22 \cite{NIST_800_22}. As presented in Table V, these data passed the test suite with a significance level of 0.01.

\begin{table}[!htb]

\centering
\caption{NIST SP800-22 test results of 1,000 datasets of 1,000,000-bit sequences collected from the final outputs. To pass the test with a significance level of 0.01, it is mandatory for each of every sub-methods that (i) the P-values-total must be above 0.0001, and (ii) the proportion must be above 0.98.} 
\vspace{0.2cm}

\begin{ruledtabular}
\begin{tabular}{lccc}

\multicolumn{1}{l}{Method}      & P-values-total & Proportion & Result \\ \cmidrule{1-4}
1. Frequency	& 0.055 361 &	0.9920 &	Pass \\
2. Block Frequency	& 0.010 165 &	0.9880 &	Pass \\
3. Runs	& 0.647 530 &	0.9870 &	Pass \\
4. Longest Run	& 0.133 404 &	0.9880 &	Pass \\
5. Rank	& 0.518 106 &	0.9850 &	Pass \\
6. Fast Fourier Transform	& 0.411 840 &	0.9940 &	Pass \\
7. Overlapping Template	& 0.233 162 &	0.9890 &	Pass \\
8. Universal	& 0.344 048 &	0.9930 &	Pass \\
9. Linear Complexity	& 0.548 314 &	0.9930 &	Pass \\
10. Approximate Entropy	& 0.846 338 &	0.9860 &	Pass \\
11. Non-overlapping Template	& 0.502 932 &	0.9896 &	Pass \\
12. Serial	& 0.317 974 &	0.9890 &	Pass \\
13. Cumulative Sums	& 0.613 197 &	0.9920 &	Pass \\
14. Random Excursions	& 0.501 651 &	0.9905 &	Pass \\
15. Random Excursions Variant	& 0.499 293 &	0.9867 &	Pass \\

\end{tabular}
\end{ruledtabular}
\end{table}

A dataset of 100,000,000-bit sequences was also submitted to an autocorrelation analysis to detect repeating patterns and evaluate the degree of similarity over a time series. The result, as plotted in Figure 9, indicates that autocorrelation values occur well below the 95\% confidence intervals for all over 100 successive bit intervals except the first data point (lag 0) that corresponds to the correlation with itself. Based on this result, it may be concluded that the final outputs have no periodic pattern over an acceptable range.

\begin{figure}[!t]
  \vspace{-5mm}
  \includegraphics[width=8.5cm,keepaspectratio]{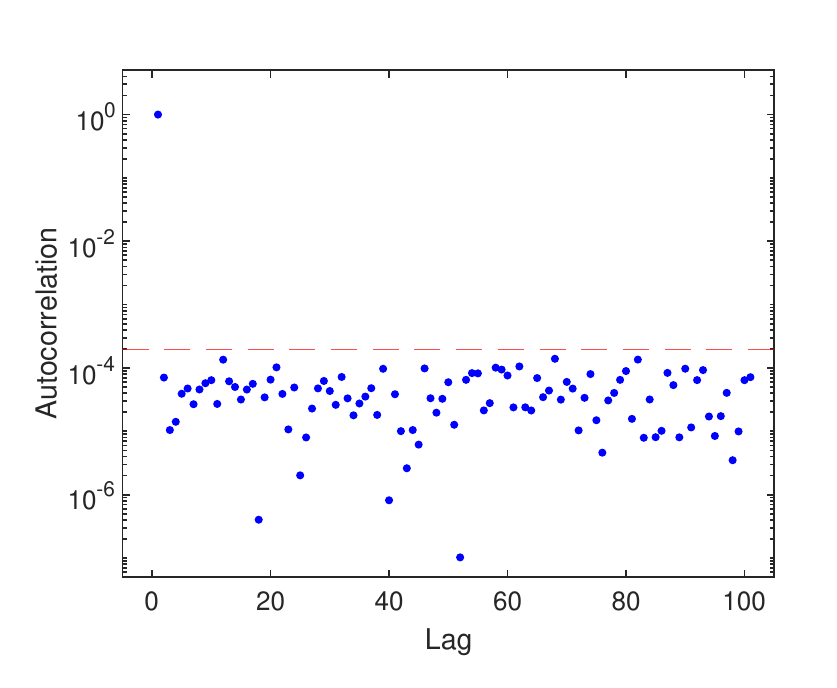}
  \vspace{-2mm}
  \caption{Autocorrelation values (solid dots) with 95\% confidence intervals (dashed line) calculated from a dataset of 100,000,000-bit sequences collected from the final outputs.}  
\end{figure}

\section{Discussion}

In this work, the experimental proof-of-concept that incorporated eight tunnel diodes for random number generation was demonstrated with the use of simple electronics. Because the preliminary implementation used an off-the-shelf microcontroller unit that has a limited time resolution, the data acquisition rate was then limited to a range of kHz. 

Ideally, it is possible to harness tunnel diodes for random number generation at low cost with high-speed capability. Appendix A presents the preliminary realization of this multiplexing scheme to achieve data acquisition at a rate of 2 MHz. Additionally, in Appendix B, we demonstrate the potential of using a tunnel diode to operate at a rate of at least 20 MHz. These demonstrations used an oscilloscope which can be replaced with the implementation of an inexpensive electronics, such as field programmable gate array, for time-counting measurement.

\section{Conclusions and outlook}

To conclude, this work demonstrated the realization of an entropy source by incorporating multiple tunnel diodes connected in series onto a single circuit. This approach significantly boosted the data rate of random number generation to make it multifold. Furthermore, raw data obtained from this multiplexing scheme were more random than those realized from the original scheme that employed a single tunnel diode. 

This paper’s preliminary implementation employed eight diodes with different I-V characteristics, and we harnessed an off-the-shelf time-counting module for data acquisition at a rate of 10 kHz. A computer was used to process bit extraction. Consequently, a sum of 64-bit data were obtained per current sweep, resulting in binary outputs that could be ideally generated at a speed of 640 kbit/s. These raw data were submitted to the Toeplitz-hashing extractor for rendering of uniformly random outputs. Final outputs were then validated with the NIST SP800-90B, showing that they achieved almost full entropy. Statistical randomness of these data passed the NIST SP800-22 test suite, and repeating patterns were not found from the autocorrelation analysis. Post-processing and statistical tests were performed with a computer.

A future improvement would be the implementation of a low-noise circuit to support high-frequency operations as well as the integration of a high-speed field programmable gate array (FPGA)\cite{FPGA, Xilin}–as a replacement of a time-counting module \cite{Zheng2017_J} for data acquisition and a replacement of a computer for bit extraction and post-processing \cite{Bai2021, Lie2020_W, Zhang2016_XG_conf, Zhang2016_XG}. This development, regarding the circuit design, as seen in Figure 4, that employs eight tunnel diodes with a maximum power consumption of 1,000 mW, may increase the capability to generate random bits, potentially up to 100 Mbit/s. To construct a random number generator that complies with industrial certification, proper engineering \cite{Balasch2018} to detect hardware failure is essential for real-world applications.
This engineering requires an integration of health-monitoring schemes in compliance with the NIST SP800-90B \cite{NIST_800_90B} or AIS 20/31 \cite{Killmann2011} (AIS: Application Notes and Interpretation of the Scheme issued by the German Federal Office for Information Security) recommendations to achieve provable robustness against security attacks.

\section*{Acknowledgements}

The authors thank Wittawat Yamwong and Apichart Intarapanich for helpful discussions as well as Sataporn Chanhorm, Siriporn Saiburee, Grit Pichayawaytin, Narusorn Doljirapisit, Jularat Nimnuan, Chadkamol Polsongkram, Pattarakon Klinhom, Parichat Muangaram and Manee Sornklin for assistance during the development of the preliminary prototype. 

This work was supported by a research program from Thailand's National Electronics and Computer Technology Center, the National Science and Technology Development Agency (NECTEC-NSTDA).


\appendix
\section{Data acquisition at a rate of 2 MHz}

This part demonstrates the capability to perform this multiplexing scheme at a higher data acquisition rate. We implemented two tunnel diodes (P/N: 3I201A and 3I201G) connected in series with the simplified flow diagram, as illustrated in Figure 10. Instead of using a differentiator circuit, a comparator circuit made of two comparators was used for detecting events of voltage jumps.

\begin{figure}[!htbp]
  \includegraphics[width=8.5cm,keepaspectratio]{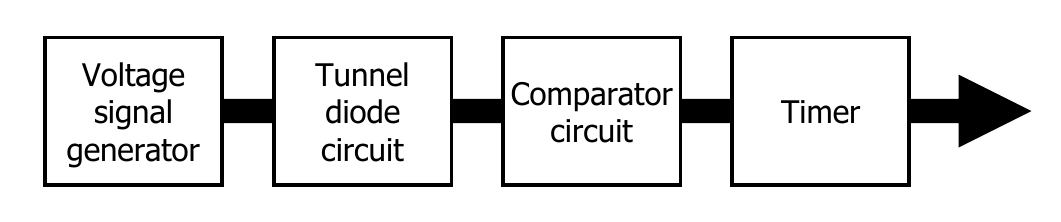}
  \vspace{-2mm}
  \caption{Flow diagram of the experiment for data acquisition at a rate of 2 MHz. A waveform generator (Agilent 33220A) was configured to produce sinusoidal voltage signals with a 2 MHz repetition rate. The voltage signals were then converted to current source used for current sweeping at the tunnel diode circuit. An oscilloscope (PicoScope 6404D) was employed as a timer to obtain values of $t_p$.}
\end{figure}

Using the circuit diagram in Figure 4, we adjusted the electronic parts to support current sweeping with a 2 MHz repetition rate. The tunnel diode circuit that employed an operational amplifier (P/N: LM833) was replaced with a faster circuit that employed a transistor (P/N: 2N3906). Implementation of the comparator circuit that used one comparator (P/N: LM311) was modified to support higher frequencies with the use of two faster comparators (P/N: MCP6561T-E/OT). An oscilloscope was employed as a timer, and linear interpolation was implemented at the events of voltage jumps to obtain values of $t_p$.

Figure 11 presents the experimental data obtained from the implementation. The time series plots were incorporated with undesired pattern, due to the defects of the implementation. These defects may have been related to the instability of the power supply and current source, which could be further circumvented with a proper circuit development.

\begin{figure}[htbp]
  \includegraphics[width=7.5cm,keepaspectratio]{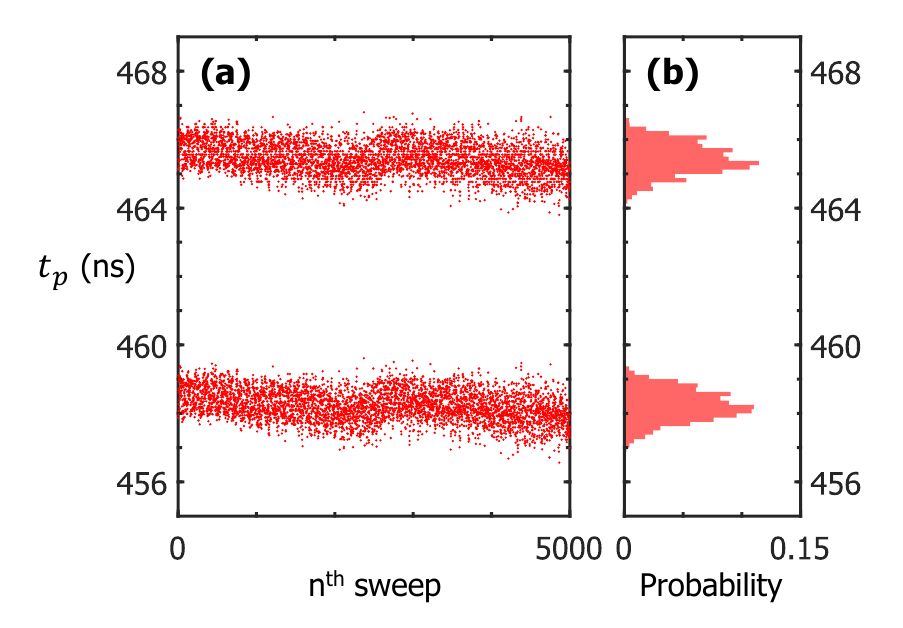}
  \vspace{-2mm}
  \caption{(a) Time series of $t_p$ values and (b) histograms of the time-series data. Current sweeping across two tunnel diodes with a 2 MHz repetition rate were implemented for the data acquisition. Data were collected from 5,000 current sweeps with a time resolution of 3.125 ps. In (b), the histogram has a bin size of 125 ns.}
\end{figure}

\begin{figure}[!h]
  \includegraphics[width=6.5cm,keepaspectratio]{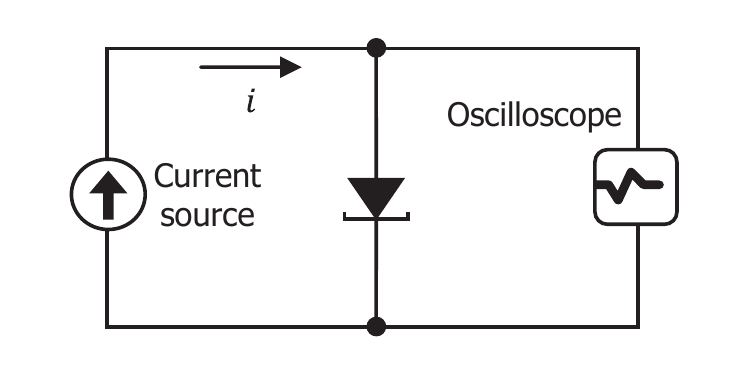}
  \vspace{-2mm}
  \caption{Circuit diagram of the experiment. A waveform generator (Agilent 33220A) was configured to produce sinusoidal voltage signals with a 20 MHz repetition rate. The voltage source was then converted to the current source for current sweeping across a tunnel diode (P/N: 3I306N). An oscilloscope (PicoScope 6404D) was employed to obtain voltage time series data. To derive values of $t_p$, voltage signals were then processed on a computer using the linear interpolation at the events of voltage jumps.}
\end{figure}

\section{Data acquisition at a rate of 20 MHz}

This part demonstrates the capability to use a tunnel diode for generating random values of $t_p$ with data acquisition at a rate of 20 MHz. This implementation followed the circuit diagram, as shown in Figure 12. A waveform generator was configured to produce a voltage source that subsequently converted to the current source for the current sweeping across a tunnel diode with a 20 MHz repetition rate. An oscilloscope was employed to collect voltage signals. Obtained data were then processed on a computer to derive values of $t_p$ as presented in Figure 13.

\begin{figure}[!h]
  \includegraphics[width=7.5cm,keepaspectratio]{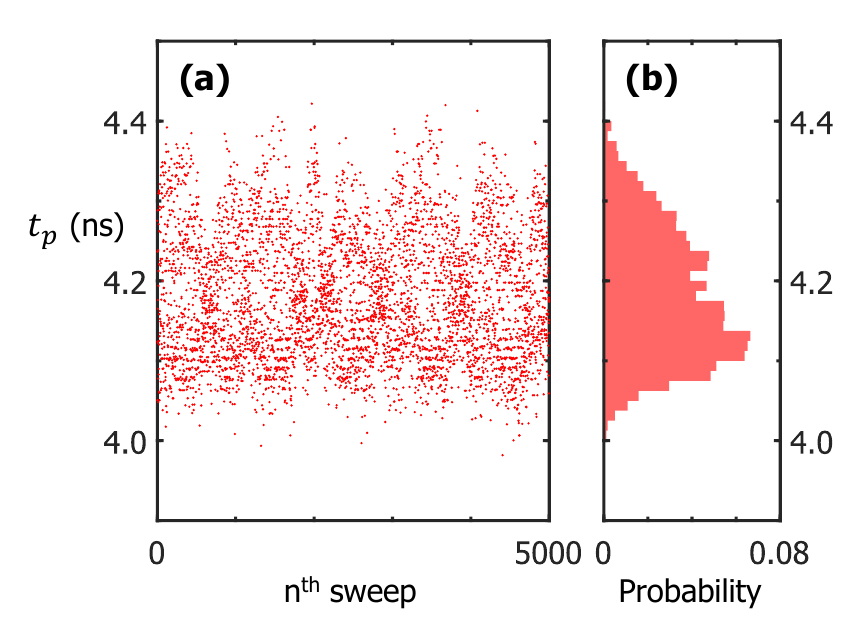}
  \vspace{-2mm}
  \caption{(a) Time series of $t_p$ values and (b) histogram of the time-series data. Current sweeping across a tunnel diode with a 20 MHz repetition rate were implemented for the data acquisition. Data were collected from 5,000 current sweeps with a time resolution of 3.125 ps. In (b), the histogram has a bin size of 12.5 ps.}
\end{figure}

\section*{References and notes}
\bibliography{mybib} 

\begin{thebibliography}{45}
\bibitem [{90B()}]{90B}%
  \BibitemOpen
  \href@noop {} {}\bibinfo {note} {Following the NIST SP800-90B
  \cite{NIST_800_90B}, 10 estimators used for non-IID evaluation are (1) the
  most common value (MCV) estimation, (2) collision estimation, (3) Markov
  estimation, (4) compression estimation, (5) t-Tuple estimation, (6) longest
  repeated substring (LRS) estimation, (7) multi most common in window
  (multiMCW) prediction estimation, (8) lag prediction estimation, (9) multi
  Markov Model with counting (multiMMC) prediction estimation, and (10) LZ78Y
  prediction estimation.}\BibitemShut {Stop}%
\end{thebibliography}

\end{document}